\documentclass[aps,showpacs,showkeys,twocolumn,groupedaddress]{revtex4}
\usepackage{amsmath,amsfonts,amssymb,mathrsfs,graphicx}

\newcommand{\ie}{{\it i.e.}}

\newcommand{\eg}{{\it e.g.}}

\newcommand{\cf}{{\it cf.}}

\newcommand{\eq}{Eq.}

\newcommand{\fig}{Fig.}

\newcommand{\Ref}{Ref.}
\newcommand{\Refs}{Refs.}

\begin{document}

\vspace*{-1.45cm}
\begin{flushright}
TUM-HEP-445/01
\end{flushright}

\title{Quark and Lepton Mass Patterns and the Absolute Neutrino Mass Scale}

\author{Manfred Lindner}
\email[E-mail address: ]{lindner@ph.tum.de}
\affiliation{Institut f{\"u}r Theoretische Physik, Physik-Department,
Technische Universit{\"a}t M{\"u}nchen (TUM), James-Franck-Stra\ss{}e,
85748 Garching bei M{\"u}nchen, Germany}

\author{Walter Winter}
\email[E-mail address: ]{wwinter@ph.tum.de}
\affiliation{Institut f{\"u}r Theoretische Physik, Physik-Department,
Technische Universit{\"a}t M{\"u}nchen (TUM), James-Franck-Stra\ss{}e,
85748 Garching bei M{\"u}nchen, Germany}

\date{\today}

\begin{abstract}
\vspace*{0.2cm}
We investigate what could be learned about the absolute
scale of neutrino masses from comparisons among the
patterns within quark and lepton mass hierarchies. First, we
observe that the existing information on neutrino masses fits quite well the
unexplained, but apparently present regularities in the quark and charged lepton
sectors. Second, we discuss several possible mass patterns, pointing out that
this is consistent with hierarchical neutrino mass patterns especially
disfavoring the vacuum solution.
\end{abstract}

\pacs{14.60.Pq, 14.60.Lm, 14.60.St}
\keywords{Absolute neutrino mass scale}

\maketitle

Non-vanishing neutrino mass squared differences imply neutrino
oscillations, which in fact have been observed in recent years. The
measurements of the mass squared splittings between the mass eigenstates
$\nu_l$, $l=1, 2, 3$, give the hierarchy $\Delta m_{21}^2 \equiv
m_2^2-m_1^2 \ll \Delta m_{32}^2 \simeq \Delta m_{31}^2$. Atmospheric neutrino
oscillations are dominated by the large mass squared splitting $\Delta
m_{32}^2 \simeq \Delta m_{31}^2 \simeq 3.3 \cdot 10^{-3}\, \mathrm{eV}^2$,
while solar neutrino oscillations allow four different solutions. These are
the so-called  LMA, SMA, LOW, and VAC solutions with 
$\Delta m_{21}^2 \simeq  3 \cdot 10^{-5} \, \mathrm{eV}^2$, 
$7 \cdot 10^{-6} \, \mathrm{eV}^2$, 
$10^{-7} \, \mathrm{eV}^2$,
$10^{-10}\, \mathrm{eV}^2$, 
respectively~\cite{Fogli:2001vr,Bahcall:2001zu,Bandyopadhyay:2001aa},
where the LMA solution is preferred after
inclusion of the latest SNO data.
So far, for the absolute neutrino mass scale only upper bounds
from several experiments exist (for an overview see, \eg,
\Ref~\cite{Law:2001sy}):  The kinematical endpoint of tritium beta decay
leads to $m_1 \leq 2.2 \, \mathrm{eV}$, while $0\nu 2\beta$-decay
(neutrinoless double beta decay) even implies a stronger bound for the
electron neutrino Majorana mass, \ie,   $m_{M}^{\nu_1} \leq 0.2 \,
\mathrm{eV}$. Furthermore, somewhat weaker but similar bounds emerge from
astrophysics and cosmology.  
However, except from these bounds, the absolute
neutrino mass scale is not yet known. Thus, in the most extreme
cases, hierarchical ($m_1 \ll | \Delta m_{21} | \equiv | m_2 - m_1 |$) or
degenerate ($m_1 \gg | \Delta m_{21} |$) mass spectra are allowed, which
ultimately should be understood in some theoretical model.
In this paper, we will observe that neutrino masses fit the well known
empirical regularities of quark and charged lepton masses. We will generalize
this discussion and use rather simple models and assumptions in order to
obtain information on the absolute neutrino mass spectrum from a
phenomenological comparison with the quark and charged lepton mass spectra.
This implies that we argue in terms of mass eigenvalues instead of mass
textures, which is an approach somewhat different from what is initiated by GUT
theories. It is however perfectly possible that some texture at the GUT scale
translates into the observed patterns. Hence, we will point into this direction 
when appropriate.

\begin{figure}[ht!] \begin{center}
\includegraphics*[width=8cm]{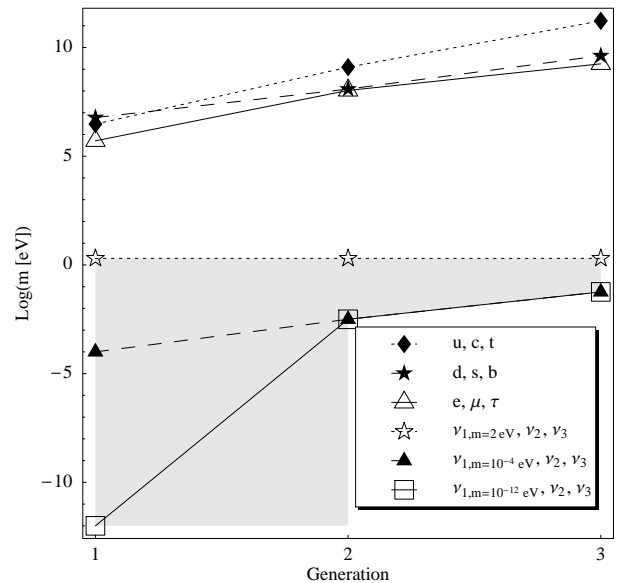}
\end{center}
\vspace*{-0.5cm}
\caption{\label{Hierarchy} Logarithm (base 10) of the quark, charged lepton,
and neutrino masses plotted over the generation number. For the quark and
charged lepton masses we choose the mean values given by
\Ref~\cite{Groom:2000in}. For the neutrino masses we assume a hierarchy
$m_1<m_2<m_3$ with the parameter values $\Delta m_{32}^2 = 3.3 \cdot 10^{-3}
\, \mathrm{eV}^2$, $\Delta m_{21}^2 = 10^{-5} \, \mathrm{eV}^2$ (LMA), as well
as different values for $m_1$. The grey-shaded region indicates the region of
allowed neutrino masses for the given LMA mass squared differences and $m_1 <
2.2 \, \mathrm{eV}$.}
\end{figure}

The regularities in the quark and charged lepton mass spectrum can be seen in
\fig~\ref{Hierarchy}, where the mean values of the lepton and quark masses
from \Ref~\cite{Groom:2000in} are plotted logarithmically over the generation
number.
The fact that the masses lie almost on straight and parallel lines suggests
that a yet unknown law describes this regularity and maybe even the ``small''
corrections to linearity. The linear behavior in this plot may, for example,
point to an exponential or power law dependence on the generation number $l$.
This immediately raises the question if similar regularities exist in the
neutrino sector and what could be learned from these.

Neutrino masses are assumed to be generated by Dirac or Majorana mass terms
in extensions of the Standard Model. In the most extreme cases, one may either
have pure Dirac masses $m_D^{\nu_l}$ or pure Majorana masses $m_M^{\nu_l}$ for
the physical neutrino masses, though mixtures between those are allowed in
general. Depending on the model, the Majorana
masses are often assumed to be created by the see-saw mechanism from Dirac
masses $m_D^l$ and heavy (right-handed) Majorana masses $M_R^l$, leading
in the absence of leptonic mixing to the see-saw mass
relation~\cite{yana79,gell79,moha80} \begin{equation} m_M^{\nu_l}
= \frac{(m_D^l)^2}{M_R^l}, \qquad l=1,2,3.
\label{seesaw}
\end{equation}
Taking into account leptonic mixing, the Dirac mass matrix $m_D$ can be
diagonalized by two unitary matrices $V_L$ and $V_R$ by
\begin{equation}
 m_D = V_L \, D \, V_R^\dagger,
 \label{mddiagonal}
\end{equation}
where $D$ is a diagonal matrix of the mass eigenvalues. Without loss of
generality, we can choose a basis where the right-handed neutrino mass matrix
$M$ is diagonal with real and positive eigenvalues. The effective neutrino mass
matrix $m_\nu$ is then given by the general see-saw
relation~\cite{Wetterich:1981bx}
\begin{eqnarray}
 m_\nu & = & - m_D \, M^{-1} \, m_D^T \nonumber \\
 & = & - \left( V_L \, D \, V_R^\dagger \right) \, M^{-1} \, \left( V_R^* \, D
\, V_L^T \right).
\label{seesawgeneral}
\end{eqnarray}
The Majorana mass eigenstates, \ie, the eigenvalues of $m_\nu$, are
then entirely determined by the term $D \, V_R^\dagger \, M^{-1} \,
V_R^* \, D$, which is a complex symmetric matrix, since unitary
transformations do not affect the eigenvalues. Thus, if $V_R^\dagger \, M^{-1}
\, V_R^*$ is approximately diagonal, \ie, $V_R \simeq 1$, and the bimaximal
mixings come from $V_L$, \eq~(\ref{seesaw}) can be used. Since we are mainly
interested in qualitative mass patterns on logarithmic scales,
\eq~(\ref{seesaw}) can also be used if $V_R$ does not change the order of
magnitudes of the eigenvalues significantly. The MNS matrix depends
in addition on a matrix $U_L^l$ diagonalizing the charged
lepton mass matrix, \ie,
\begin{equation}
U_{MNS} = \left( U_L^l \right)^\dagger \, \tilde{V}_L,
\label{mns}
\end{equation}
where $\tilde{V}_L$ is diagonalizing $m_\nu$ in \eq~(\ref{seesawgeneral}).
Thus, in a basis where the charged lepton sector is diagonal, $U_{MNS}
=\tilde{V}_L$. Note, however, that from $U_{MNS}$ no direct information on $V_R$
can be obtained, which means that mixings can enter \eq~(\ref{seesawgeneral}),
which are depending on the neutrino mass model. We will further on initially
focus on \eq~(\ref{seesaw}), which we will especially use as a tool in order to
obtain small neutrino masses but not as a key ingredient of our mechanisms. Then
we will discuss what this could imply for a realistic see-saw model.

We define a mass ordering $m_1 < m_2 < m_3$, \ie,
the mixing  angles are defined correspondingly and the relations
\begin{equation}
 m_i \ge \Delta m_{ij} \equiv m_i - m_j
 \label{ordering}
\end{equation}
with $i,j = 1,2,3$ and $i>j$ are satisfied in general.
Thus, for given mass squared differences the
neutrino mass spectrum is fixed, except from the absolute scale given by
the absolute mass of one of the neutrinos. In
\fig~\ref{Hierarchy}, we added the information on the neutrino masses by
plotting the curves for the largest allowed mass $m_1 \simeq 2 \,
\mathrm{eV}$ and two smaller values for $m_1$. Comparing the neutrino masses
for different absolute mass scales $m_1$ with the charged lepton and  quark
mass hierarchies, we make the following interesting observations:
\renewcommand{\labelenumi}{(\arabic{enumi})}
\begin{enumerate}
\item
Equation~(\ref{ordering}) implies that for given mass squared
differences the values of $m_3$ and $m_2$ are bounded from below, \ie,
for $m_1 \rightarrow 0$ $m_3 \rightarrow \Delta m_{32}$ and $m_2
\rightarrow \Delta m_{21}$. This leads to the the grey-shaded region in
\fig~\ref{Hierarchy}, which is the region of all
allowed mass spectra for the LMA solution used in the figure. The smallest
values of $m_2$ and $m_3$ thus determine the steepest slope for the neutrino
mass values between the generations two and three. Comparing this slope with
the corresponding slopes of the charged leptons and quarks shows that they are
apparently parallel. This observation suggests that there may be connections
between the regularities of quark or lepton masses and neutrino masses. In
addition, it points towards a hierarchical mass ordering, \ie, $m_1 \le \Delta
m_{21}$.
\item
In \fig~\ref{Hierarchy}, all of the quark
and charged lepton masses approximately lie on a straight line. One may
assume that there is a  theoretical reason for that and may thus
expect the same  regularity for the neutrino masses. This would imply that the
neutrino  masses also lie on an approximately straight line, parallel to
one of the hierarchies of the other masses. Depending on what
reference hierarchy is chosen, it would fix the absolute
neutrino mass scale to $m_1 \simeq 10^{-5\pm 2} \, \mathrm{eV}$, as well as it
is consistent with the LMA solution.
\item
The left-handed quarks and leptons can be symmetrically arranged in
electroweak doublets. One might therefore expect that the splittings
of quark and lepton masses are somehow correlated to their electroweak isospin
properties, which may, for example, imply that the isospin splittings of the quarks in
\fig~\ref{Hierarchy} are related to the isospin splittings of
the leptons. Note, however, that the absolute neutrino mass scale is, compared
to their isospin  $+1/2$ quark equivalents, shifted down by a large unknown
quantity.  This shifting is often believed to be done by the heavy
Majorana masses $M_R$ introduced in the see-saw mechanism in
\eq~(\ref{seesaw}).   
\end{enumerate}

\begin{table*}
\caption{\label{TSummary} The different schemes as well as their
results. The values for $m_2$ and $m_3$ for the scheme (A-2) are
only rough estimates. The value of $\Delta m_{32}^2$ was in all cases fixed to
be $\Delta m_{32}^2 = 3.3 \cdot 10^{-3} \, \mathrm{eV}^2$. The value of $\Delta
m_{21}^2$ was given in the scheme (A-2) and derived in all
other schemes.}
\begin{center}
\begin{ruledtabular}
\begin{tabular}{lcccccc}
Scheme & $m_1 \, [\mathrm{eV}]$ & $m_2 \, [\mathrm{eV}]$ & $m_3 \,
[\mathrm{eV}]$ & $\Delta m_{21}^2 \,
[\mathrm{eV^2}]$ & $\Delta m_{32}^2 \, [\mathrm{eV^2}]$ & Favored solution \\
\hline
(A-2) Rough curve comparison & $10^{-7} - 10^{-3}$ & $ \sim 10^{-3} -
10^{-2}$ & $\sim 10^{-2} - 10^{-1}$ & $1.0 \cdot 10^{-5}$ & $3.3 \cdot
10^{-3}$ & LMA/SMA/LOW \\ \hline
(B-2a) Reference curve selected (a)& $1.7 \cdot
10^{-5}$ & $3.4 \cdot 10^{-3}$ & $5.8 \cdot 10^{-2}$ & $1.2 \cdot 10^{-5}$ &
$3.3 \cdot 10^{-3}$ & LMA \\   
(B-2b) Reference curve selected (b)& $4.8 \cdot 10^{-9}$ & $2.0 \cdot
10^{-4}$ & $5.7 \cdot 10^{-2}$ & $4.2 \cdot 10^{-8}$ & $3.3 \cdot 10^{-3}$ &
LOW \\  (B-3) Isospin symmetry used & $2.1 \cdot 10^{-7}$ & $8.7 \cdot
10^{-4}$ & $5.7 \cdot 10^{-2}$ & $7.6 \cdot 10^{-7}$ & $3.3 \cdot 10^{-3}$ &
SMA/LOW \\  
\end{tabular}
\end{ruledtabular}
\end{center}
\end{table*}

The above observations suggest that there exist empirical meaningful
mass relations which indeed may allow to deduce the absolute neutrino mass
scale. There are, however, different ways to combine the existing information
such that different possibilities emerge. Before we will discuss
some of them, let us approach the problem from a different
point of view. If we assume all of the $\Delta m^2$'s to be known from
measurements, we will only have to find values for one unknown parameter determining the
absolute mass scale, such as $m_1$.
However, without any {\em a priori} information, we could also fit
two or three of the unknown parameters $m_1$, $\Delta m_{21}^2$, and $\Delta
m_{32}^2$ to their equivalents of the charged lepton
and quark curves. Thus, we may distinguish three cases:
\renewcommand{\labelenumi}{(\Alph{enumi})} \begin{enumerate}
\item
We assume $\Delta m_{21}^2$ as well as $\Delta m_{32}^2$ to be known from
measurements. Then the absolute mass scale $m_1$ can be chosen such that the
slopes of the neutrino mass curve approximately fit the slopes of one of the
reference mass curves. As noted above, this leads to
$m_1 \simeq 10^{-7} - 10^{-3} \, \mathrm{eV}$, 
depending on what hierarchy we use for reference. Thus, without
additional assumptions we will not obtain preciser information.  We label
this case (A-2) for linking option (A) with observation (2).
\item
We assume only one of the $\Delta m^2$'s to be known and two parameter values
have to be found. We choose $\Delta m_{32}^2$, because it is better
established and measured without ambiguities.  Using
additional assumptions about the selection of the reference curve (\cf,
observation (2)) or about the electroweak isospin symmetry (\cf, observation
(3)), we can then calculate the absolute values for the masses as well as
$\Delta m_{21}^2$. The small mass squared splitting can then be used for
comparison of the result with the possible solutions LMA, SMA, LOW,
and VAC.
\item
We assume none of the $\Delta m^2$'s to be known. This will
not provide any information on the absolute neutrino mass scale.
\end{enumerate}

We have seen that only option (B) has the potential to predict
specific numerical values for the absolute neutrino masses.
The simplest case is linking option (B) with observation (2), (B-2),
which means that we choose a specific mass hierarchy for reference.
We may, for example, assume that the physical neutrino masses are Dirac masses
and directly proportional to their charged lepton partners, \ie,
\begin{equation}
 m_D^{\nu_l} = \tilde{C} \cdot m_D^l,
 \label{Dirac}
\end{equation}
where $\tilde{C}$ is some generation number-independent constant.
This might also point towards a connection between the lepton masses
within each electroweak isospin doublet.
The constant $\tilde{C}$ can now be determined by
the measured value of $\Delta m_{32}^2 = 3.3 \cdot 10^{-3} \, \mathrm{eV}^2$
from \eq~(\ref{absm}) in order to find $\tilde{C} \simeq 3.2 \cdot
10^{-11}$.  We can then calculate the other mass squared differences $\Delta
m_{31}^2 \simeq \Delta m_{32}^2$ and $\Delta m_{21}^2 \simeq 1.2 \cdot 10^{-5}
\, \mathrm{eV}^2$ in fairly good agreement with the LMA solution. For the
absolute masses we finally obtain from \eq~(\ref{Dirac}) $m_1 \simeq 1.7
\cdot 10^{-5} \, \mathrm{eV}$, $m_2 \simeq 3.4 \cdot 10^{-3} \, \mathrm{eV}$,
and $m_3 \simeq 5.8 \cdot 10^{-2} \, \mathrm{eV}$, referred to as case (B-2a),
which is in perfect agreement with the well-known constraints to neutrino
masses. In this case, however, the smallness of $\tilde{C}$ is not very
appealing and often believed to be achieved by the see-saw mechanism, such as
in \eq~(\ref{seesaw}) with the charged lepton masses for the Dirac masses
$m_D^l$ in this equation. For example, we may assume that the physical
neutrino masses are Majorana masses and, for some
reason, the heavy right-handed Majorana masses follow the same hierarchy as the
the charged lepton masses,
\ie,
\begin{equation} M_R^l  = K \cdot m_D^l,
\label{Right} \end{equation}
where $K$ is a generation number independent constant.
Then we obtain for the Majorana neutrino masses from \eq~(\ref{seesaw})
\begin{equation}
 m_M^{\nu_l} = \frac{(m_D^l)^2}{M_R^l} =
\frac{m_D^l}{K}.  \label{absm}
\end{equation}
Now we immediately see the connection between the case of physical Dirac
neutrino masses in \eq~(\ref{Dirac}) and the case of physical Majorana neutrino
masses in \eq~(\ref{absm}): they are mathematically equivalent for $K =
1/\tilde{C}$. It is obvious from \fig~(\ref{Hierarchy}) that $\tilde{C}$ has to
be very small to shift the absolute neutrino mass scale down from the charged
lepton scale. This fine-tuning is often assumed to be done by the see-saw
mechanism introducing the large Majorana mass scale, such as done in
\eq~(\ref{Right}) here. Of course, \eq~(\ref{seesaw}) does not include leptonic
mixings, which means that one could ask what one could learn about the mixings
in the general see-saw case in \eq~(\ref{seesawgeneral}). It is obvious from
the discussion there that this see-saw mechanism would be consistent for $V_R$
almost diagonal, \ie, close to unity. This is an assumption quite often used
in texture models, such as in \Refs~\cite{King:1998jw,Shafi:2000su,King:2001uz}.
In addition, other matrices $V_R$ not affecting the order of magnitudes of the
eigenvalues of $m_\nu$ could be thought about, since we consider logarithmic
scales. However, our discussion does not apply to cases when the eigenvalue
structure is radically changed by $V_R$.

Another possibility is that only the Dirac neutrino masses are related to the
charged lepton masses, since these masses are produced by the same type of
Yukawa couplings. Assuming the right-handed heavy Majorana mass to be
universal, \ie, generation index independent, we can write
\begin{equation}
 m_M^{\nu_l} = \frac{(m_D^{\nu_l})^2}{M_R} \quad \mathrm{with} \quad
m_D^{\nu_l} =  R \cdot m_D^l.
\label{case2b}
\end{equation}
Since in this case the right-handed Majorana mass matrix $M$ in
\eq~(\ref{seesawgeneral}) commutes with $V_R$, the combination $V_R^ \dagger \,
V_R^*$ gives the unit matrix at least in the absence of CP violation, which
means that \eq~(\ref{case2b}) is quite general. Using the same procedure
as above, we obtain in this case (B-2b) for the Majorana neutrino masses $\Delta
m_{21}^2 \simeq 4.2 \cdot 10^{-8} \, \mathrm{eV}^2$, $m_1 \simeq 4.8 \cdot
10^{-9} \, \mathrm{eV}$, $m_2 \simeq 2.0 \cdot 10^{-4} \, \mathrm{eV}$, and $m_3
\simeq 5.7 \cdot 10^{-2} \, \mathrm{eV}$. Note that here we do not relate the
physical neutrino masses to quark or charged lepton masses and therefore do not
have parallel curves.

Instead of choosing some specific mass hierarchy for reference, we may
use the electroweak isospin argument from observation (3) in order to
create a case (B-3). Assuming that the weak isospin $I = \pm 1/2$
lepton masses follow the same scheme as the $I = \pm 1/2$ quark masses and
ignoring lepton mixings, we could postulate that
\begin{equation}
\frac{m_D^{\nu_l}}{m_D^l} = C \cdot
\frac{m_D^{l,I=+1/2}}{m_D^{l,I=-1/2}}
\label{scale}
\end{equation}
with $C$ a generation independent constant and $m_D^{l,I= \pm 1/2}$ the Dirac
quark masses of generation $l$ with weak isospin $I= \pm 1/2$. We can
now again calculate the constant $C$, for instance, from the value of $\Delta
m_{32}^2 \simeq 3.3 \cdot 10^{-3} \, \mathrm{eV}^2$, obtaining $C \simeq 8.1
\cdot 10^{-13}$. Then the values of $\Delta m_{31}^2$ and $\Delta m_{21}^2$ are
determined by \eq~(\ref{scale}) and can be evaluated to be $\Delta m_{31}^2
\simeq \Delta m_{32}^2$ and $\Delta m_{21}^2 \simeq 7.6 \cdot 10^{-7} \,
\mathrm{eV}^2$. For the absolute neutrino masses we here obtain $m_1 \simeq 2.1
\cdot 10^{-7} \, \mathrm{eV}$, $m_2 \simeq 8.7 \cdot 10^{-4} \, \mathrm{eV}$,
and $m_3 \simeq 5.7 \cdot 10^{-2} \, \mathrm{eV}$.
Note, however, that the weak isospin is actually defined in terms of flavor
eigenstates and not mass eigenstates which we are using here. Taking into
account mixings, linear combinations of neutrino masses would enter at least
in $m_D^{\nu_l}$ in \eq~(\ref{scale}). Thus, the plausibility of this approach
would strongly constrain lepton mixings or point more towards a generation index
number rather than isospin dependent property.

\begin{figure}[ht!]
\begin{center}
\includegraphics*[width=8cm]{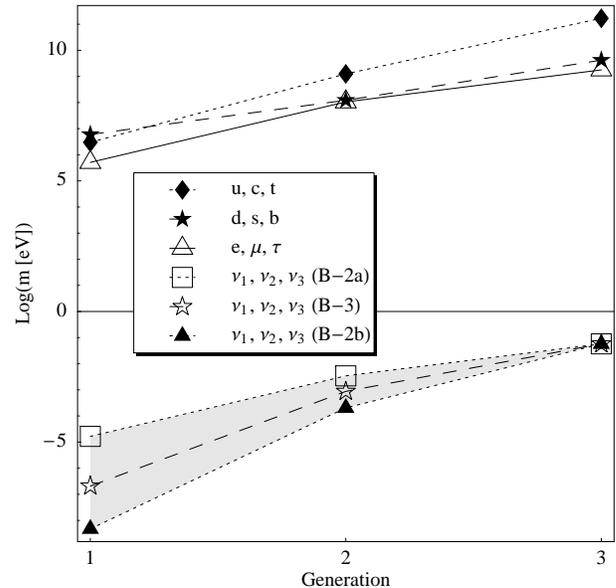}
\end{center}
\vspace*{-0.5cm}
\caption{\label{Calculated} Logarithm (base 10) of the quark, charged lepton,
and neutrino masses obtained from our calculations plotted over the
generation number. For the quark and charged lepton masses we chose the mean
values given by \Ref~\cite{Groom:2000in}. For the neutrino masses we assumed
 a hierarchy $m_1<m_2<m_3$ with the parameter value $\Delta
m_{32}^2 = 3.3 \cdot 10^{-3} \, \mathrm{eV}^2$. We obtained for the small
mass squared splittings $\Delta m_{21}^2 = 1.2 \cdot 10^{-5} \,
\mathrm{eV}^2$ (B-2a), $\Delta m_{21}^2 = 7.6 \cdot 10^{-7} \, \mathrm{eV}^2$
(B-3), and $\Delta m_{21}^2 = 4.2 \cdot 10^{-8} \, \mathrm{eV}^2$ (B-2b). The
grey-shaded region gives an estimate for possible mass spectra from
assumptions similar to the ones in this paper.}
\end{figure}
Let us come back
to our original argument in observation (2), \ie, that we want to have a
neutrino mass spectrum with slopes such that it looks similar to the charged
lepton and quark curves. Figure~\ref{Calculated}, showing the results of our
three calculations, indicates that the scheme (B-2a) fits best the other
quark and charged lepton mass hierarchies. This scheme predicted a $\Delta
m_{21}^2 = 1.2 \cdot 10^{-5} \, \mathrm{eV}^2$, which is in agreement with the
LMA solution. Since we were only using very simple models and only one possible
value for $\Delta m_{32}^2$, some factor difference from the measured value does
not destroy this conclusion.

In summary, we presented very simple, purely phenomenological
approaches to extract absolute values for the neutrino masses. The mass
patterns which have been assumed are essentially power laws, which may arise
in models when neutrino masses are generated radiatively, \eg, in
Froggatt-Nielsen-like models~\cite{Froggatt:1979nt}. By comparing the
physical neutrino mass curve plotted over the generation number with the
charged lepton or quark mass curves indicated that $m_1 \simeq 10^{-7} -
10^{-3} \, \mathrm{eV}$, fixing the absolute neutrino mass scale for known
mass squared differences and their signs. Using different assumptions
together with simple Dirac and Majorana neutrino mass models allowed us to
extract numerical values for the absolute neutrino masses in several models. As
a starting point, we used the knowledge about the neutrino mass squared
differences to extract and validate these models. Table~\ref{TSummary}
summarizes the results from our calculations. In general, from such an approach
one expects a neutrino mass spectrum between the curves of the cases (B-2a)
(with linear scaling in the charged lepton masses) and (B-2b) (quadratic scaling
in the charged lepton masses), indicated by the grey-shaded region in
\fig~\ref{Calculated}. All values obtained for the absolute masses agree with
the well-known constraints to neutrino masses and follow, in fact, rather
similar patterns. One of the most important results of such patterns would be
that $m_3 \simeq 5.8 \cdot 10^{-2} \, \mathrm{eV}$ in all cases consistent with
a hierarchical neutrino mass spectrum, \ie, $m_1$ is quite small compared to the
mass differences. In addition, from these purely empirical investigations the
LMA solution provided the most appealing results by comparing the physical mass
curves of charged lepton and neutrino masses. Moreover, the VAC solution did not
fit any of our estimates. Even though such an empirical approach is by far no
theory of neutrino masses, it may point to the right absolute neutrino mass
scale by using the yet unexplained fermion mass patterns, \ie, the patterns of
mass eigenstates. We have also commented on the implications of mixings and
neutrino mass models. With this method we obtained absolute
neutrino masses in good agreement with all constraints, such that it could be
regarded as a hint for the absolute neutrino masses. Finally, it should be
interesting to investigate how the patterns in the mass eigenstates are related
to textures in the mass matrices.

We would like to thank S. Antusch, M. Freund, P. Huber, J. Kersten, T. Ohlsson,
M. Ratz, and especially G. Seidl for useful discussions and comments.
This work was supported by the ``Studienstiftung des deutschen Volkes'' (German
National Merit Foundation) [W.W.], and the ``Sonderforschungsbereich 375
f{\"u}r Astro-Teilchenphysik der Deutschen Forschungsgemeinschaft''.

\end{document}